\documentstyle[12pt]{article}

\newcommand{\matel}[3]{\langle #1|#2|#3\rangle}
\newcommand{\ra}{\rightarrow}

\newcommand{\sG}{\sigma \cdot G}

\newcommand{\aver}[1]{\langle #1\rangle}

\newcommand{\MeV}{\,\mbox{MeV}}

\newlength{\dinwidth}
\newlength{\dinmargin}
\begin{document}
{}~~\\
\vspace{1cm}
\begin{flushright}
UND-HEP-95-BIG08\\
July 1995\\
\end{flushright}

\begin{center}
\begin{Large}
\begin{bf}
%
%
QUO VADIS, FASCINUM?
\footnote{Invited Lecture given at the Workshop on the
Tau/Charm Factory, Argonne National Laboratory, June 21 -
23, 1995}
\\
\end{bf}
\end{Large}
\vspace{5mm}
\begin{large}
%
%
I.I. Bigi\\
\end{large}
%
%
Physics Dept., University of Notre Dame du Lac,
Notre Dame, IN 46556, U.S.A.\\
e-mail address: BIGI@UNDHEP.HEP.ND.EDU
\vspace{5mm}
\end{center}
\noindent
\begin{abstract}
The recent progress in our understanding of heavy-flavour decays
allows us to define more reliably which future measurements on
charm decays are
needed to further advance our understanding of QCD and to get our
theoretical  tools ready for treating beauty decays.
After sketching the theoretical landscape I list those required measurements.
I argue that some -- in particular those concerning
inclusive semileptonic charm decays and their lepton spectra --
can presumably be performed only at a tau-charm factory.

\end{abstract}
%
\section{Overview}

There are four questions I would like to address in reviewing the
long-term goals of charm decay studies:
{\em why, how, where and when?}

{\em Why?}

\noindent There exists a triple motivation for a detailed
analysis of charm decays: (i) it provides novel probes of QCD,
(ii) sharpens our tools for dealing with beauty
decays and (iii) represents one of the more promising avenues to
finding the hoped-for `unexpected', namely the intervention of New
Physics; this would most clearly be realized
through the observation of CP asymmetries
\cite{BURDMAN,HQ94}.

{\em How?}

\noindent A comprehensive program is required with its
cornerstones being the measurements of (a) lifetimes,
(b) semileptonic branching ratios,
(c) nonleptonic branching ratios and
(d) lepton spectra in exclusive as well as
inclusive semileptonic charm decays.

{\em Where?}

\noindent Three different experimental environments
have been considered
for high-statistics charm decay studies, namely
$(\alpha )$ $B$ factories, $(\beta )$ photo- and hadro-production
and $(\gamma )$ a $\tau$-charm factory.

{\em When?}

\noindent There are two benchmark dates in evaluating the
impact of a $\tau$-charm factory, namely

\noindent (A) 1998 - 2000: the next round of fixed target
experiments at FNAL will be finished by then; the data on
charm decays from CLEO will have reached a new dimension statistically
and systematically; BABAR and BELLE will commence data taking;

\noindent (B) $\sim$ 2005: the analysis of charm and beauty
decays at the $B$ factories should have reached a fully
mature level; new initiatives like CHARM2000 will have had a run;
the LHC with its dedicated program on beauty physics will hopefully
start up.

My talk will be organized as follows: in Sect.2 I will sketch the
relevant theoretical framework for charm decays; in Sect.3 I list the
database that is required -- or at least desired -- for a deeper
understanding to come about; in Sect.4 I review various
experimental stages before presenting an outlook in Sect.5.

\section{Theoretical Framework}

\subsection{The Landscape}

Four second-generation theoretical technologies have emerged
for treating the impact of the strong interactions on heavy-flavour
decays:

\noindent $\bullet$ {\em QCD Sum Rules} are employed for
describing exclusive semileptonic and non-leptonic
decays of hadrons with strangeness, charm and beauty.

\noindent $\bullet$ {\em Lattice Simulations of QCD} have
worked their way down in distance scale to deal with
exclusive semileptonic and non-leptonic
charm decays, albeit only in the quenched approximation;
quite significant jumps are required before beauty decays
can be tackled and one can go beyond
the quenched approximation.

\noindent $\bullet$ {\em Heavy Quark Effective
Theory} (HQET) on the other hand deals with exclusive
semileptonic decays of beauty hadrons; its applicability
to charm decays is dubious.

\noindent $\bullet$ {\em $1/m_Q$ Expansions} are distinct
from the other technologies in that they deal with
inclusive transitions, semileptonic as well as non-leptonic ones.
Similarly to HQET they apply best to beauty decays while
their numerical usefulness in charm decays is a priori uncertain.

There is one point of particular relevance for our discussion here.
Charm decays occupy a special place in the theoretical landscape:
they represent common ground for all the QCD methods listed
above, albeit sometimes only at the extreme range of their
applicability \footnote{The non-perturbative contributions can be
expressed through an expansion in powers of
$\mu _{had}/m_Q$; for
charm one has $\mu _{had}/m_Q \sim 0.4$. It is at least smaller than
unity, but not by a large factor.}. Comparing the predictions
from the various theoretical technologies, which emphasize
complementary aspects of QCD, with detailed and comprehensive
data has a two-fold benefit:

\noindent $\bullet$ It provides valuable insights into the inner
workings of QCD in general, and on the interplay between
perturbative and non-perturbative effects in particular.

\noindent $\bullet$ These findings can be extrapolated to
beauty decays and applied there with
{\em tested} confidence.

\subsection{Inclusive Transitions}

\subsubsection{Total Rates}

Total rates for a heavy-flavour hadron $H_Q$ to decay
into an inclusive final state $f$ can be expressed
through an expansion in powers of $1/m_Q$ \cite{SV}; through order
$1/m_Q^3$ one obtains the following master equation\cite{BUV}:
$$\Gamma (H_Q\ra f)=\frac{G_F^2m_Q^5}{192\pi ^3}|KM|^2
\left[ c_3^f\matel{H_Q}{\bar QQ}{H_Q}+
c_5^f\frac{
\matel{H_Q}{\bar Qi\sG Q}{H_Q}}{m_Q^2}+ \right.
$$
$$\left. +\sum _i c_{6,i}^f\frac{\matel{H_Q}
{(\bar Q\Gamma _iq)(\bar q\Gamma _iQ)}{H_Q}}
{m_Q^3} + {\cal O}(1/m_Q^4)\right]  \eqno(1)$$
where the dimensionless coefficients $c_i^f$ depend on the
parton level
characteristics of $f$ (such as the ratios of the final-state quark
masses
to $m_Q$); $KM$ denotes the appropriate combination of KM
parameters,
and $\sG = \sigma _{\mu \nu}G_{\mu \nu}$
with $G_{\mu \nu}$ being the gluonic field strength tensor. The last
term
in eq.(1)
implies also the summation over the four-fermion operators with
different light flavours $q$.

It is through the quantities
$\matel{H_Q}{O_i}{H_Q}$ that the dependence on the {\em decaying
hadron} $H_Q$, and
on
non-perturbative forces in general, enters; they reflect the
fact that the weak decay of the heavy quark $Q$ does not proceed
in empty space, but within a cloud of light degrees of
freedom -- (anti)quarks and gluons -- with which $Q$ and
its decay products can interact strongly.  The practical usefulness
of the $1/m_Q$ expansion rests on our ability to determine the
size of these matrix elements.
Through order $1/m_Q^3$ there are three types of expectation
values that determine the non-perturbative corrections:

\noindent (i) The leading term can be expanded further:
$$\matel{H_Q}{\bar QQ}{H_Q} = 1-
\frac{\aver{(\vec p_Q)^2}_{H_Q}}{2m_Q^2}+
\frac{\aver{\mu _G^2}_{H_Q}}{2m_Q^2} + {\cal O}(1/m_Q^3)
\eqno(2)$$
where $\aver{(\vec p_Q)^2}_{H_Q}\equiv
\matel{H_Q}{\bar Q(i\vec D)^2Q}{H_Q}$ denotes the average
kinetic energy of the quark $Q$ moving inside the hadron
$H_Q$ and $\aver{\mu _G^2}_{H_Q}\equiv
\matel{H_Q}{\bar Q\frac{i}{2}\sG Q}{H_Q}$. The first term on the
right-hand-side of eq.(2) represents the naive
spectator ansatz.

\noindent (ii) The quantity $\aver{\mu _G^2}_{P_Q}$ is known
from the meson hyperfine splitting:
$$ \aver{\mu _G^2}_{P_Q}\simeq
\frac{3}{4} (M_{V_Q}^2-M_{P_Q}^2)\; , \eqno(3)$$
where $P_Q$ and $V_Q$ denote the pseudoscalar and vector mesons,
respectively.  Therefore
$$\aver{\mu _G^2}_B\simeq 0.37\, GeV \; \;  , \; \;
\aver{\mu _G^2}_D\simeq 0.41\, GeV \eqno(4a)$$
For $\Lambda _Q$ and $\Xi _Q$ baryons one has
instead
$$\aver{\mu _G^2}_{\Lambda _Q, \Xi _Q} \simeq 0 \eqno(4b)$$
since the light diquark system inside $\Lambda _Q$ and $\Xi _Q$
carries no spin, whereas
$$\aver{\mu _G^2}_{\Omega _Q} \neq 0 \eqno(4c)$$
For $\aver{(\vec p_Q)^2}_{H_Q}$
there exists an estimate from a QCD
sum rules analysis\cite{QCDSR}
$$\aver{(\vec p_b)^2}_B
\simeq 0.5\, \pm \, 0.1\, GeV \eqno(5a)$$
and one can expect one from
lattice QCD in the foreseeable future. We do have a
model-independant lower
bound on it\cite{OPTICAL}:
$$\aver{(\vec p_b)^2}_B \geq
 0.37\, \pm \, 0.1\, GeV \; .\eqno(5b)$$
The
{\em difference} in the kinetic energy of Q inside
baryons and mesons can be related to the masses of charm and
beauty hadrons:
$$ \langle (\vec p_Q)^2\rangle _{\Lambda _Q}-
\langle (\vec p_Q)^2\rangle _{P_Q} \simeq
\frac{2m_bm_c}{m_b-m_c}\cdot
\{ [\langle M_B\rangle -M_{\Lambda _b}]-
[\langle M_D\rangle -M_{\Lambda _c}] \}
\eqno(5c)$$
where $\langle M_{B,D}\rangle$ denote the `spin averaged' meson
masses: $ \langle M_B\rangle \equiv \frac{1}{4}(M_B+3M_{B^*})$
and likewise for $\langle M_D\rangle$.
Using data
one finds: $\langle (\vec p_Q)^2\rangle _{\Lambda _Q}-
\langle (\vec p_Q)^2\rangle _{P_Q}= -(0.07 \pm 0.20)(GeV)^2$;
i.e., the present measurement of $M_{\Lambda _b}$
is not yet sufficiently accurate, but this will change
in the foreseeable future.

\noindent (iii) The expectation values for the four-quark operators
taken between {\em meson} states can be expressed in terms of a
single quantity, namely the decay constant:
$$\matel{H_Q(p)}{\bar Q_L \gamma _{\mu}q_L)
(\bar q_L \gamma _{\nu}Q_L)}{H_Q(p)}\simeq
\frac{1}{4} f^2_{H_Q}p_{\mu}p_{\nu} \eqno(6)$$
where factorization has been assumed. The theoretical
expectations for the decay constants are \cite{GUIDO}
$$f_D \simeq 200\, \pm 30\, MeV\; \; \; , \; \; \;
f_B \simeq 180 \, \pm 30\, MeV \eqno(7a)$$
$$f_{D_s}/f_D \simeq 1.15 - 1.2\; \; \; , \; \; \;
f_{B_s}/f_B \simeq 1.15 - 1.2 \eqno(7b)$$
The size of the expectation values taken between
{\em baryonic} states are quite uncertain at present. There
exists more than one relevant contraction, and for the time
being quark model estimates provide us with the only guidance!
I will return to this point when discussing predictions of
baryon lifetimes.

To illustrate the method I give the semileptonic and
non-leptonic widths for charm hadrons through order $1/m_c^2$:
$$\Gamma _{SL}(H_c) \simeq \Gamma _0
\matel{H_c}{\bar cc}{H_c}\cdot
\left( 1- \frac{2\aver {\mu _G^2}_{H_c}}{m_c^2} +
{\cal O}(1/m_c^3)\right) \eqno(8a)$$
$$\Gamma _{NL}(H_c) \simeq \Gamma _0 N_C
\matel{H_c}{\bar cc}{H_c}\cdot \left[ A_0
\left( 1- \frac{2\aver {\mu _G^2}_{H_c}}{m_c^2}\right) -
4A_2 \frac{2\aver {\mu _G^2}_{H_c}}{m_c^2} +
{\cal O}(1/m_c^3)\right] \eqno(8b)$$
$$\Gamma _0 = \frac{G_F^2m_c^5}{192\pi ^3} |V(cs)|^2\; ,
\eqno(8c)$$
where $A_{0,2}$ denote perturbative QCD corrections; I have
ignored here the small phase space correction due to
$m_s^2/m_c^2 \neq 0$. The following results can be read off
from eqs.(8):

\noindent $\bullet$ For $m_c \ra \infty$ the parton model spectator
expression is recovered.

\noindent $\bullet$ $\Gamma _{SL}$ as well as $\Gamma _{NL}$ receive
non-perturbative corrections, as does $BR_{SL}(H_c)$; the latter
quantity is lowered since the last term in eq.(8b)
enhances $\Gamma _{NL}$ ($A_2 < 0$ !)\cite{BUV}.

\noindent $\bullet$ $\Gamma _{SL}$ is {\em not} universal for all
charm hadrons $H_c$ once ${\cal O}(1/m_c^2)$ contributions
are included. One actually finds
$$\Gamma _{SL}(D)\; / \; \Gamma _{SL}(\Lambda _c) \; / \;
\Gamma _{SL}(\Omega _c) \; \sim \; 1\; / \; 1.5 \; / \; 1.2 \; ,
\eqno(9)$$
where the difference in the $\Lambda _c$ and $\Omega _c$
widths to this order are due to eqs.(4b,c);
i.e., the ratio of semileptonic branching ratios does
{\em not} coincide
with the ratio of lifetimes when comparing mesons and
baryons.

\noindent $\bullet$ The widths for $D$ and $\Lambda _c$ decays
differ already in order $1/m_c^2$.

\noindent $\bullet$ The differentiation between the widths for
$D^0$, $D^+$ and $D_s$ mesons occurs at order $1/m_c^3$ as
expressed in eq.(1), but left out in eq.(8). \footnote{$SU(3)_{Fl}$
symmetry is very much observed in the expectation values
$\aver{\mu _G^2}_{D,D_s}$ and $\aver{(\vec p_c^2)}_{D,D_s}$.}

The underlying pattern can be expressed as follows:
$$\Gamma (\Lambda _Q) = \Gamma (P_Q^0) = \Gamma (P_Q^+)
+ {\cal O}(1/m_Q^2) \eqno(10a)$$
$$\Gamma (\Lambda _Q) > \Gamma (P_Q^0) \simeq \Gamma (P_Q^+)
+ {\cal O}(1/m_Q^3) \eqno(10b)$$
$$\Gamma (\Lambda _Q) > \Gamma (P_Q^0) > \Gamma (P_Q^+)
+ {\cal O}(1/m_Q^4) \eqno(10c)$$
i.e., one predicts $\tau (\Lambda _b) < \tau (B_d) < \tau (B^-)$
as well as $\tau (\Lambda _c) < \tau (D^0) < \tau (D^+)$.
Yet keeping in mind that the $1/m_Q$ expansion is at
best a semi-quantitative tool
for $m_Q = m_c$ ($\mu _{had}/m_c \sim 0.4$!) one cannot
expect to make precise predictions for the charm lifetime ratios.
On the other hand -- and that is a central point of my message --
measurements of the lifetime ratios for all weakly decaying
charm hadrons can be used with great profit to disentangle
the various contributions in charm decays. Such an anatomy
will then pave the way for more reliable predictions on beauty
decays. To say it differently: a comprehensive study of charm decays
can be harnessed as Nature's microscope onto the
numerically smaller effects in beauty decays.

In Tables \ref{TABLE1} and \ref{TABLE2}
I juxtapose the theoretical expectations and
predictions on charm and beauty lifetime ratios
\cite{MIRAGE,DS,STONE2,BARYONS1,REPORT} with present
data \cite{MALVEZZI,SHARMA}.
\begin{table}
\begin{tabular} {|l|l|l|}
\hline
Observable &QCD ($1/m_c$ expansion) &Data \\
\hline
\hline
$\tau (D^+)/\tau (D^0)$ & $\sim 2 \; \; $
 [for $f_D \simeq$ 200 MeV]
&$2.547 \pm 0.043$ \\
&(mainly due to {\em destructive} interference) & \\
\hline
$\tau (D_s)/\tau (D^0)$ &$1\pm$ few$\times 0.01$
&  $ 1.125\pm 0.042$ \\
\hline
$\tau (\Lambda _c)/\tau (D^0)$&$\sim 0.5\; ^*$ &
$0.51\pm 0.05$\\
\hline
$\tau (\Xi ^+ _c)/\tau (\Lambda _c)$&$\sim 1.3\; ^*$ &
$1.75\pm 0.36$\\
\hline
$\tau (\Xi ^+ _c)/\tau (\Xi ^0 _c)$&$\sim 2.8\; ^*$ &
$3.57\pm 0.91$\\
\hline
$\tau (\Xi ^+ _c)/\tau (\Omega _c)$&$\sim 4\; ^* $&
$3.9 \pm 1.7$\\
\hline
\end{tabular}
\centering
\caption{QCD Predictions for Charm Lifetimes}
\label{TABLE1}
\end{table}
\begin{table}
\begin{tabular} {|l|l|l|}
\hline
Observable &QCD ($1/m_b$ expansion) &Data \\
\hline
\hline
$\tau (B^-)/\tau (B_d)$ & $1+
0.05(f_B/200\, \MeV )^2
[1\pm {\cal O}(10\%)]>1$ &$1.04 \pm 0.05$ \\
&(mainly due to {\em destructive} interference) & \\
\hline
$\bar \tau (B_s)/\tau (B_d)$ &$1\pm {\cal O}(0.01)$
&  $ 0.98\pm 0.08$ \\
\hline
$\tau (\Lambda _b)/\tau (B_d)$&$\sim 0.9\; ^*$ & $0.76\pm 0.06$
\\
\hline
\end{tabular}
\centering
\caption{QCD Predictions for Beauty Lifetimes}
\label{TABLE2}
\end{table}
As mentioned before, at present one has to rely on quark models
to estimate the size of the relevant {\em baryonic} expectation values.
Thus there is a model dependance in the predictions on
{\em baryon} lifetimes -- in contrast to the case with meson
lifetimes. This is indicated in the tables by an asterisk.

The general agreement with the data is remarkable, in particular
for the charm system, where the expansion parameter
is not much smaller than unity. A few more detailed comments are in order:

\noindent $\bullet$ The $D^+$-$D^0$ lifetime difference is driven
mainly by a destructive interference \cite{PI} with `Weak
Annihilation' (WA) contributing not more than 10 - 20\%.
Within the accuracy of the expansion, the
data are reproduced.

\noindent $\bullet$ The $D_s$-$D^0$ lifetime ratio can be treated
with better theoretical accuracy, namely of order a few percent.
The observed near equality of $\tau (D_s)$ and $\tau (D^0)$
represents rather direct evidence for the reduced weight of WA in
charm meson decays\cite{DS}.

\noindent $\bullet$ Even the expectations on the charm baryon
lifetimes reproduce the data which is quite remarkable since there
are constructive as well as destructive contributions to baryon
lifetimes. It has to be noted though that the present measurements
suffer from large uncertainties.

\noindent $\bullet$ However the prediction on
$\tau (\Lambda _b)/\tau (B_d)$ appears to be in
serious (though not yet conclusive) disagreement with the data. The
details of what went into that prediction can be found in
ref.\cite{REPORT}; here I want to state only the following conclusion.
If
$\tau (B_d)$ indeed exceeds $\tau (\Lambda _b)$ by 25 - 30 \%,
then a `theoretical price' has to be paid. It strongly suggests that
the present agreement between theoretical
expectations and data on charm baryon lifetimes is largely
accidental and most likely would not survive in the face of more
precise measurements! To state it in a more constructive
manner: more precise measurements on charm baryon
lifetimes would then allow to isolate the source of the discrepancy
between prediction and observation.

As already said, on general grounds one does not predict the
semileptonic widths to be the same for all charm hadrons --
apart from $\Gamma _{SL}(D^0)\simeq \Gamma _{SL}(D^+)$,
which is protected by isospin invariance and Cabibbo suppression.
A priori there could be a sizeable difference in
$\Gamma _{SL}(D^0)$ vs. $\Gamma _{SL}(D_s)$ due to a
WA contribution to the latter observable. It is then a non-trivial
prediction that those two quantities largely coincide:
$$ 1\pm \sim \, {\rm few}\, \% = \frac{\tau (D_s)}{\tau (D^0)}
\simeq \frac{BR_{SL}(D_s)}{BR_{SL}(D^0)} \simeq
1\pm \sim 10\% \eqno(11)$$
On the other hand a sizeable difference is expected in
the semileptonic widths of baryons and mesons which is
expressed as follows:
$$BR_{SL}(\Lambda _c) > BR_{SL}(D^0)\cdot
\frac{\tau (\Lambda _c)}{\tau (D^0)} \simeq
0.5 \cdot BR_{SL}(D^0) \eqno(12)$$

\subsubsection{Lepton Spectra}

A detailed study of the lepton spectra in inclusive semileptonic
decays of $D^0$, $D^+$ and $D_s$ mesons is highly desirable.
One expects \cite{WA} sizeable differences between the
energy spectra in $D^0$ and $D_s$  and to a lesser degree also in
$D^+$ decays. For there is a WA process that
is Cabibbo allowed [forbidden] for $D_s$ [$D^+$] mesons where
the hadrons in the final state emerge from (double) gluon
emission of the initial anti-quark line. These differences will
show up mainly in the endpoint region. An analogous
complication is expected for semileptonic $B$ decays:
hadronization affects the spectra in
the endpoint region differently in
$B_d$ and $B^-$ transitions. This creates  a systematic uncertainty
in the value extracted from inclusive decays that cannot
be evaluated reliably unless

\noindent $\bullet$ one separates $B_d$ and $B^-$ decays or

\noindent $\bullet$ measures the corresponding effects for $D$ mesons
and extrapolates to $B$ mesons through a $1/m_Q$ expansion.

\subsection{Exclusive Charm Decays}
\subsubsection{Leptonic and Semileptonic Channels}

Measuring $BR(D^+,D_s \ra \mu \nu ,\, \tau \nu )$ with
{\em good} accuracy represents a high priority goal since it
allows to extract the decay constants $f_D$ and $f_{D_s}$.
There exists considerable intrinsic interest in the value of these
hadronic parameters; in addition -- and maybe more
urgently from a phenomenological perspective -- once
$f_D$ and $f_{D_s}$ have been well measured, one can
confidently extrapolate to the beauty sector and predict
$f_B$ and $f_{B_s}$.

As discussed in detail in El'Khadra's talk
\cite{AIDA} a host of theoretical
tools can be brought to bear on exclusive semileptonic
charm decays: QCD sum rules, Lattice QCD and HQET in addition
to quark models. Confronting their predictions with
comprehensive measurements of the relevant hadronic
form factors and their dependance on the momentum transfer
will provide us with valuable insights into the inner
workings of QCD; it also will be of great benefit in quantitatively
understanding exclusive semileptonic $B$ decays.

\subsubsection{Nonleptonic Modes}

Most important is a general caveat: the relationship
between the pattern in exclusive modes and in inclusive
transitions is quite tenuous. The former in contrast to the
latter are very sensitive to the dramatic behaviour of QCD
in the infrared regime; there exist relatively straightforward
examples \cite{MIRAGE} showing that while {\em individual}
exclusive rates get enhanced or decreased significantly
by the strong
interactions, these effects average out to a large degree
in the sum.  No
theoretical tools have been developed yet that can master
these complexities and at present one can employ only models
of uncertain reliability. Nevertheless there is a valid motivation
behind such `phenomenological engineering', in particular
when applied to two-body modes. For it allows us -- once
sufficiently many branching ratios have been well measured --
to extract the size of the contributing isospin amplitudes and
their phase shifts \cite{BUCELLA}. This provides information
that is essential for
designing a strategy for CP studies and for interpreting
its outcome.

\subsubsection{Radiative Decays}

While in the Standard Model there is no (short-distance)
penguin operator generating $D\ra \gamma + X,\,
\gamma K^*,\, \gamma \rho/\omega $ transitions, long distance
dynamics can. One should note that even the inclusive rate
receives contributions from a non-local (though
higher-dimensional)  operator. Thus the radiative branching
ratios cannot be predicted in a reliable fashion. Yet measuring
$BR(D\ra \gamma K^*,\, \gamma \rho /\omega)$ helps us
in two ways:
On the one hand one can again extrapolate
to the beauty system and obtain a reliable estimate for the
impact of long-distance dynamics on the corresponding modes
$B \ra \gamma K^*,\, \gamma \rho /\omega $. This is important
for any attempt to extract $|V(ub)|$ from these radiative $B$
decays. On the other hand one has opened up a new
window onto New Physics; for it can manifest itself by
producing a significant deviation of the ratio
$BR(D\ra \gamma \rho /\omega )/BR(D\ra \gamma K^*)$
from $\tan ^2(\theta _c)$.

\section{Required/Desired Database}

The preceding discussion should have made it clear that
even without aiming at possible manifestations of New
Physics the need for further data on charm decays has
not diminished; the advances in our
theoretical understanding actually allow us to define
more precisely the kind of future measurement one needs
for further progress.  I will briefly sketch them.

While there is no need from theory to measure the $D^+$, $D^0$ and
$\Lambda _c$ lifetimes more precisely, it would be quite
useful to determine $\tau (D_s)$ to within 1\%. Clearly the
greatest need for improvement exists for  the $\Xi _c$ and
$\Omega _c$ lifetimes. A 5-10\% accuracy in $\tau (\Xi _c^0)$,
$\tau (\Xi _c^+)$ and $\tau (\Omega _c)$  would enable us to
extract the size of the relevant baryonic matrix elements.

The benchmark to aim for in $BR(D^+,D_s \ra \mu \nu , \tau \nu )$
is a 10\% accuracy allowing to extract the decay constants to within
5\%.

For practical reasons one wants to know the {\em absolute}
branching ratios for charm hadrons to within a few percent.
Such information which is sorely missing for $D_s$, $\Lambda _c$
and $\Xi _c$ decays
\cite{ROUDEAU} is needed, among other things, for
proper charm counting in $B$ decays and for determining the
absolute values of $BR(B_s \ra l \nu D_s^{(*)})$,
$BR(B \ra D \bar D_s^{(*)})$ and $BR(B_s \ra D_s^{(*)} \bar D_s^{(*)})$.
Likewise one wants to know the inclusive rates for
$\Lambda _c \ra \Xi +X_s$, $\Xi _c \ra \Xi +X$ etc. Our
ignorance here constitutes a major bottle neck in
many studies, like using $l\Xi$ correlations to
distinguish between $\Lambda _b$ and $\Xi _b$ decays.

It is also important to know the absolute branching ratios for the
inclusive transitions $D_s \ra l+X$, $\Lambda _c \ra l +X$,
$\Xi _c \ra l +X$ to complement the information obtained from
the lifetimes.   A detailed analysis of the lepton spectra in
$D,D_s \ra l + X$ (and also in $\Lambda _c \ra l+X$) would be
of great theoretical help when extracting $|V(ub)|$ from the
endpoint region in $B\ra l +X$ decays.

The dependance of the hadronic form factors in
exclusive semileptonic charm decays  on the momentum transfer
has to be measured directly and the analysis has to be extended
to include also channels like $D^+, D_s \ra l \nu \eta / \eta ^{\prime}$.

The data base for the program of `theoretical engineering' in two-body modes
referred to above has to be completed by analysing final states
containing (multi)neutrals.

Enough statistics has to be accumulated to study doubly Cabibbo
suppressed decays in detail.

The radiative channels $D,D_s \ra \gamma K^*, \gamma \rho / \omega$
have to be searched for in a dedicated manner.

Finally it would be quite useful to remeasure the reaction
$e^+ e^- \ra D \bar{D} +X$ for $E_{c.m.} \sim 5-6$ GeV. Old
SPEAR data suggest an enhancement there; if true, it would point
to rather virulent final state interactions in that interval.
That region happpens to be the one that is also probed in
$B\ra D \bar D_{(s)}$; such effects would have an obvious impact
on the CP phenomenology in those $B$ decays.

\section{Experimental Stage}

The measurements listed above fall into three categories:

\noindent (A) The lifetimes can be measured by fixed target
experiments (or at $B$ factories).

\noindent (B) Some of the measurements might not be
impossible in hadronic collisions or at a $B$ factory,
but certainly represent a stiff challenge
there \cite{WISS}. Determining $BR(D^+,D_s \ra \mu \nu )$
with a 20\% accuracy
presumably belongs into that category, as do observing
non-leptonic decays with multi-neutrals in the final state and
extracting
{\em absolute} branching ratios for $D_s$ mesons and the charmed baryons.
A $\tau$-charm factory on the other hand offers the cleanest
measurements \cite{ROUDEAU}.

\noindent (C) There are finally measurements that
presumably will
remain in the sole domain of a $\tau$-charm factory. Among them
are: studies of the lepton spectra in $D/D_s/\Lambda _c \ra l + X$;
the semileptonic branching ratios for the various charm hadrons
and the identification of genuine
radiative $D$ decays which requires the efficient rejection of
nonleptonic modes like $D\ra K^* \pi ^0 \ra K^*\gamma [\gamma ]$, i.e.
where one photon escapes detection; this can probably be achieved
only by making use of the excellent energy resolution available due
to beam-energy constraints at a threshold machine.

\section{Outlook}

Let me start out with some very general statements which I then
relate to the purpose of our meeting. A comprehensive program on
Heavy Flavour Physics is essential in any serious quest to unveil
Nature's Grand Design. Detailed studies of charm decays have to form
an integral part of such a program. In an `ideal' or at least
`optimal' world a $\tau$-charm factory plays a central role in such
studies. This factory is justified by its unique capabilities to advance
our understanding of QCD,
and it would run with good
luminosity in the energy range 3 GeV $\leq \sqrt{s} \leq$ 6 GeV;
its discovery potential for New Physics
would serve as the `icing on the cake'.
The `ideal' world is defined as one where a $\tau$-charm factory would
be running by now; in an `optimal' world it would start delivering data
by the end of the millenium, like the asymmetric $B$ factories.
Not surprisingly, our world is not ideal and probably not optimal.
There is still an excellent motivation for a first-class
$\tau$-charm factory starting up later, yet the emphasize will shift
somewhat. High precision studies of $\tau$ and charmonium physics
would represent the superb primary justification. On the other
hand, the battle lines for open charm physics might be re-drawn.
I expect that our experimental colleagues will devise some ingenious
new methods for obtaining at least decent measurements of
absolute branching ratios for charmed baryons etc. For cost and
time (also running time) reasons it might make more sense then to
limit the effective energy range of the machine to
3 GeV $\leq \sqrt{s} \leq  $ 4.4 GeV and
concentrate on the core part of open charm physics, namely the
weak decays of $D$ and $D_s$ mesons with a two-fold
purpose: to
perform measurements of absolute branching ratios, semileptonic
decays and their lepton spectra that cannot be
made in other experimental environments
-- and to follow up on tantalizing hints
for the intervention of New Physics that might have surfaced in the
meantime!

{\bf Acknowledgements:} I am grateful to T.D. Lee for sharing
his insights and his enthusiasm with us. It was a
stimulating meeting nicely organized by J. Repond. This work was
supported by the National Science Foundation under
grant number PHY 92-13313.  I also thank the Institute for
Nuclear Theory at the University of Washington for its
hospitality and the Department of Energy for partial
support during the write-up of this manuscript.

\end{document}